\documentclass[12pt]{article}

\usepackage[height=8.85in,width=6.45in]{geometry}

\usepackage[utf8]{inputenc}
\usepackage{amsmath}
\usepackage{amssymb}
\usepackage{bm}
\usepackage{mathtools}
\numberwithin{equation}{section}
\usepackage{slashed}
\usepackage{braket}
\usepackage{hyperref}
\usepackage{cite}
\usepackage{xcolor}
\usepackage{tikz}
\usepackage{tikz-cd}
\usepackage{times}
\usepackage{courier}
\usepackage[boxsize=0.5em,aligntableaux=center]{ytableau}

\usepackage{xcolor}
\usepackage{mdframed}

\renewenvironment{figure}[1][]{
  \begin{originalfigure}[#1]
    \begin{mdframed}[linecolor=black!0,backgroundcolor=black!2]
}{
    \end{mdframed}
  \end{originalfigure}
}

\def\bZ{\mathbb{Z}}
\def\fsu{\mathfrak{su}}
\def\fso{\mathfrak{so}}

\def\fsp{\mathfrak{sp}}
\def\fe{\mathfrak{e}}
\def\ff{\mathfrak{f}}
\def\fg{\mathfrak{g}}
\def\fu{\mathfrak{u}}

\def\Nequals#1{$\mathcal{N}{=}#1$}

\newcommand{\vev}[1]{ \left\langle {#1} \right\rangle }

\DeclareMathOperator{\Tr}{Tr}
\DeclareMathOperator{\tr}{tr}

\usepackage{slashed}

\def\CA{{\cal A}}

\def\CC{{\cal C}}
\def\CD{{\cal D}}

\def\CM{{\cal M}}

\def\BR{{\mathbb R}}

\def\BZ{{\mathbb Z}}

\def\U{\mathrm{U}}
\def\SU{\mathrm{SU}}

\def\SO{\mathrm{SO}}
\def\Sp{\mathrm{Sp}}

\def\Spin{\mathrm{Spin}}

\def\so{\mathfrak{so}}

\def\g{\mathfrak{g}}

\def\d{{\rm d}}

\def\beq#1\eeq{\begin{align}#1\end{align}}

\def\PD{\mathop{\mathrm{PD}}}
\def\Sym{\mathop{\mathrm{Sym}}\nolimits}

\newcommand{\asymm}{\ydiagram{1,1}}
\newcommand{\fund}{\ydiagram{1}}

\begin{document}

\begin{titlepage}

\begin{flushright}
IPMU-17-0139\\
MPP-2017-220
\end{flushright}

\vskip 3cm

\begin{center}

{\Large \bfseries 8d gauge anomalies and the topological\\[.5em]
 Green-Schwarz mechanism}

\vskip 1cm
I\~naki Garc\'ia-Etxebarria$^1$,
Hirotaka Hayashi$^2$,
Kantaro Ohmori$^3$,\\[.5em]
Yuji Tachikawa$^4$, and 
Kazuya Yonekura$^4$
\vskip 1cm

\begin{tabular}{ll}
1. & Max Planck Institute for Physics, \\
& F\"ohringer Ring 6, 80805 Munich, Germany\\
2. & Department of Physics, School of Science, Tokai University,\\
& 4-1-1 Kitakaname, Hiratsuka-shi, Kanagawa 259-1292, Japan  \\
3. & School of Natural Sciences, Institute for Advanced Study, \\
& 1 Einstein Drive Princeton, NJ 08540, USA\\
4.  & Kavli Institute for the Physics and Mathematics of the Universe, \\
& University of Tokyo,  Kashiwa, Chiba 277-8583, Japan
\end{tabular}

\vskip 1cm

\end{center}

\noindent
String theory provides us with 8d supersymmetric gauge theory with
gauge algebras $\fsu(N)$, $\fso(2N)$, $\fsp(N)$, $\fe_{6}$,
$\fe_{7}$ and $\fe_{8}$, but no construction for $\fso(2N{+}1)$,
$\ff_4$ and $\fg_2$ is known.  In this paper, we show that the
theories for $\ff_4$ and $\fso(2N{+}1)$ have a global gauge anomaly
associated to $\pi_{d=8}$, while $\fg_2$ does not have it. 
We argue that the anomaly associated to $\pi_d$ in $d$-dimensional gauge theories
cannot be canceled by topological degrees of freedom in general.  We
also show that the theories for $\fsp(N)$ have a subtler gauge
anomaly, which we suggest should be canceled by a topological analogue
of the Green-Schwarz mechanism.
\end{titlepage}

\tableofcontents

\section{Introduction and Summary}
In ten dimensions with \Nequals1 supersymmetry, no supersymmetric
gauge theory is anomaly free, whereas with the supergravity multiplet
only the gauge algebras $\fso(32)$ and $\fe_8\oplus \fe_8$ are
allowed. Both possibilities are realized by string theory
\cite{Green:1984sg,Gross:1984dd,Adams:2010zy}.
 \footnote{An argument
   based purely on anomalies of the relevant supersymmetry multiplets
   would allow for the additional possibilities
   $\fe_8\oplus \fu(1)^{248}$ and $\fu(1)^{496}$, but more careful
   consideration of the Green-Schwarz couplings required by
   supersymmetry rules these theories out \cite{Adams:2010zy}.}
Furthermore, in these models, the anomaly of the fermions is canceled
by a special coupling of the gauge field and the metric to the
two-form field, now known as the Green-Schwarz mechanism \cite{Green:1984sg}.

We study a question of a similar nature in eight dimensions.\footnote{There are a series of works in six dimensions initiated by Kumar and Taylor \cite{Kumar:2009us}. An elegant review of the eight dimensional models with gravity and their construction in string theory can be found in \cite{Taylor:2011wt}. }
A vector multiplet in eight dimensional \Nequals 1 supersymmetry contains a gaugino which is a chiral fermion, and is specified by a gauge algebra.
A question is then which algebra is anomaly-free, and which case is actually realized in string theory.

As is well-known, the $\fsu(N)$ theory is realized on $N$ D7-branes,
and the $\fso(2N)$ theory with $N\ge 4$ and $\fsp(N)$ theory are
realized on $N$ D7-branes on top of an O7$^\mp$-plane.  In addition,
the F-theory 7-branes provide the $\fe_{6,7,8}$ theories
\cite{Vafa:1996xn,Morrison:1996na,Morrison:1996pp,Gaberdiel:1997ud}.
What then is the status for the other gauge algebras $\fso(2N{+}1)$,
$\ff_4$ and $\fg_2$?  Is it simply that we do not know the
construction yet, or are they inconsistent because of an
anomaly?\footnote{In perturbative type IIB string theory it might appear that it is possible to
  engineer the $\fso(2N{+}1)$ theory by putting an odd number of D7
  branes on top of an O7$^-$ plane. There are various arguments that this
  configuration is inconsistent: there is no
  homology group or K-theory class that could support the discrete RR torsion associated
  with the stuck D7 brane \cite{Hyakutake:2000mr,Bergman:2001rp}, the would-be
  monodromy of the axio-dilaton around the stack does not correspond
  to any element in the Kodaira classification, and a D3 probe of the
  configuration would have a 4d $\SU(2)$ anomaly in its worldvolume
  \cite{Hyakutake:2000mr}. The analysis in this paper shows that the
  perturbative IIB construction is indeed 
  inconsistent, both by a direct computation of the anomaly in 8d and
  by refining the D3 probe argument.}

In this paper, we will first show that the cases $\fso(2N{+}1)$ with $2N{+}1\ge 7$ and $\ff_4$ have global anomalies in the sense of 
Witten \cite{Witten:1982fp,Witten:1985xe} associated to $\pi_8$ of respective groups.
This will be done in two ways, one by using the traditional method of Elitzur and Nair \cite{Elitzur:1984kr}, another by introducing a BPST instanton.
The second method reveals a surprise: although the case $\fsp(N)$ does not have a global anomaly associated to $\pi_8$, it has in fact a subtler anomaly detected by the $\eta$ invariant as discussed e.g.~in \cite{Witten:2015aba,Witten:2016cio}.

This begs the question how this is consistent with the fact that the D7-branes on an O7$^+$-plane give rise to an eight-dimensional (8d) supersymmetric $\fsp(N)$ gauge theory.
We suggest that this is via a coupling to topological degrees of freedom which cancel the anomaly,
in a way analogous to the Green-Schwarz mechanism,
although we have been unable to actually describe the topological quantum field theory (TQFT) which does the job.
Morally speaking, it should be given by the topological part of the Ramond-Ramond (RR) fields in the spacetime with an O7$^+$-plane, but the authors have been unable to write it down.
Instead we provide a simple three-dimensional (3d) model where the gauge anomaly is canceled by a TQFT.

This also begs an independent question whether the traditional global anomaly associated to $\pi_d$ in $d$-dimensions might be canceled by coupling to a TQFT. 
Although the answer to this question is already implicit in the original arguments of global anomalies~\cite{Witten:1982fp,Witten:1985xe}, 
we will make it more explicit that this is impossible, settling the non-existence of the 8d theories with gauge algebras $\so(2N+1)$ and $\ff_4$.

Intriguingly, we do not find any anomaly for $\fg_2$ in this paper.
It is not known whether this algebra can be realized in some
compactification of string theory down to eight dimensions.

The rest of the paper is organized as follows: In
Sec.~\ref{sec:review}, we start by reviewing the global anomaly
associated to $\pi_d$ and its relation to the mod-2 index and the
$\eta$ invariant.  In Sec.~\ref{sec:trad}, we then compute the
anomalies of 8d \Nequals1 gauge theories associated to $\pi_{d=8}$
using the traditional method.  In Sec.~\ref{sec:new}, we re-compute
the same anomalies using a new method using the instanton
background. We reproduce the results of Sec.~\ref{sec:trad} and also
find a subtler anomaly for $\fsp(N)$.  Sections~\ref{sec:trad} and
\ref{sec:new} can be read independently.  We then show in
Sec.~\ref{sec:can't} that the traditional global anomaly associated to
$\pi_d$ cannot be canceled by coupling it to a TQFT.  Finally in
Sec.~\ref{sec:topGS} we discuss how a subtler global anomaly can
sometimes be canceled by a TQFT in a way analogous to the standard
Green-Schwarz mechanism.  We have an appendix listing $\pi_{d\le 11}$
of various Lie groups.

\section{Brief review of traditional global anomaly}
\label{sec:review}
In this section, we review the global anomaly associated to $\pi_d$ 
and its relation to the mod-2 index and the $\eta$ invariant.
This section is completely standard and can be skipped by an experienced reader.
\subsection{Traditional anomaly associated to \texorpdfstring{$\pi_d$}{homotopy group}}
\label{sec:pid.anomaly}
First let us briefly review how the traditional anomaly arises. 
Gauge transformations $g$ on $\BR^d$ which go to 1 at infinity are classified topologically by $\pi_d(G)$,
so we will loosely write it as $g \in \pi_d(G)$. (More precise statement is that the equivalence class $[g]$ of such gauge transformations under continuous deformation
is classified by $\pi_d(G)$, $[g] \in \pi_d(G)$.)
Then we consider a family of gauge fields $a(t)$ parametrized by $t \in \BR$ given by
\beq
a(t) = f(t) g^{-1}dg,
\eeq
where $f(t)$ is a smooth function of $t$ such that $f(t {\to}{ -}\infty) =0 $ and $f( t{ \to} {+}\infty) = 1$.
Now, consider the fermion partition function $\det P_L\slashed{D}[a(t)]$ in the background field $a(t)$, where $P_L$ is the projector to left-handed fermions.
The gauge field configurations at $t \to -\infty$ and $t \to +\infty$ are gauge equivalent.
However, it can happen that 
\beq
\det P_L\slashed{D}[a(t = +\infty)] = e^{i\CA} \det P_L\slashed{D}[a(t = -\infty)].
\eeq
The phase $e^{i \CA}$ represents the anomaly.

In a situation in which $\det P_L\slashed{D}[a(t)]$ is naturally real,
as in Witten's four-dimensional (4d) $\Sp(N)$ anomalies~\cite{Witten:1982fp} as well as 8d anomalies of
strictly real representations, the anomaly takes values in
$e^{i\CA} = \pm 1$ and it is simply related to the number of
eigenvalues of the Dirac operator which cross zero when $t$ is
changed from $-\infty$ to $+\infty$. Moreover, by spectral flow
considerations, this number can be represented by the number of zero
modes in $d+1$ dimensions. We take a gauge field $A$ on $\BR^{d+1}$
which is just $a(t)$ by regarding $t$ as one of the directions of
$\BR^{d+1}$. Then we can compute the number of zero modes $J$ of the
Dirac operator on $\BR^{d+1}$ in the presence of the gauge field
$A$. The anomaly is simply given by $e^{i \CA} = (-1)^J$.

In summary, the traditional anomaly for real $\det P_L\slashed{D}[a(t)]$ is detected by considering a gauge field configuration in one-higher dimension $d+1$.
We take a gauge field $A$ on $\BR^{d+1}$ which goes to a pure gauge at infinity $A \to g^{-1} dg$, where $g$
corresponds to a nontrivial element of $\pi_d(G)$ at the sphere $S^d$ at infinity. 
Then, if the number of fermion zero modes are odd, we have the traditional global anomaly.

\subsection{Mod-2 index and the \texorpdfstring{$\eta$}{eta} invariant}

Before moving on, we also review the so-called mod-2 index of Dirac operators to see the relation between the traditional anomaly~\cite{Witten:1982fp} reviewed above
and a more modern proposal in \cite{Witten:2015aba}.
The mod-2 index can be defined when the Dirac operator (times the imaginary number $i$) can be taken
to be real-antisymmetric. This means the following.  The Dirac
operator is $\CD = i \gamma^\mu D_\mu$. Here
$D_\mu=\partial_\mu +T_a A^a_\mu$ is a covariant derivative, where $T_a$
is the representation of generators of the gauge algebra $\g$, and for
simplicity we suppress the spin connection. Then, the reality of the Dirac
operator means that $\gamma_\mu \otimes 1$ and $1 \otimes T_a$ are the
strict-real representation of ${\rm Clif} \oplus \g$, where
${\rm Clif}$ is the Clifford algebra associated to the tangent
bundle of a space-time manifold. This is possible if the representations of both of ${\rm Clif}$ and $\g$ are
strict-real, or both of them are pseudo-real.\footnote{The strict and pseudo real conditions are rephrased as the existence of a charge conjugation matrix $C$
such that $C^T= C $ for strict real and $C^T=-C$ for pseudo real, respectively. 
The charge conjugation matrix of a representation of the algebra ${\rm Clif} \oplus \g$ is given by $C_{\rm Clif} \otimes C_\g$ in an obvious notation. 
Hence it is strictly real if both of ${\rm Clif} $ and $ \g$ are strict/pseudo real.}
Witten's $\Sp(N)$ anomaly in 4d
uses the case where they are both pseudo-real. In our application in
this paper, the gauginos are in the adjoint representation which is
strict-real, and also the 9-dimensional Clifford algebra is strictly
real (in the convention $\{ \gamma_\mu, \gamma_\nu \}=2\delta_{\mu \nu}$).

Now we can define the mod-2 index. First, recall the case of finite dimensional matrices.
A finite dimensional real anti-symmetric matrix $A$ can be transformed by an orthogonal matrix to a block diagonal form as
\beq
A \to (0)  \oplus  \cdots  \oplus (0) \oplus 
\left( \begin{array}{cc} 0 & \lambda_1 \\ -\lambda_1 & 0 \end{array} \right)
\oplus \cdots \oplus 
\left( \begin{array}{cc} 0 & \lambda_\ell \\ -\lambda_\ell & 0 \end{array} \right) \label{eq:antimatrix}
\eeq
where each $(*)$ is one of the blocks of the block-diagonal matrix. Then, the number of zero eigenvalues modulo 2 does not change under 
smooth continuous deformations. 

On a closed spin manifold, the operator $\gamma^\mu D_\mu$ can be
regarded as an infinite-dimensional anti-symmetric real matrix.  This
is because $\gamma_\mu \otimes 1$ are real-symmetric matrices,
$1 \otimes T_a$ are real-antisymmetric, and the derivatives
$\partial_\mu$ are real-antisymmetric (by integration by
parts on a closed manifold). Therefore, the number of zero eigenvalues mod 2 is well
defined. This is the mod-2 index ${\rm Ind}(\CD) \in \BZ_2$ of the
Dirac operator $\CD = i \gamma^\mu D_\mu$.
By using this notation, the global anomaly discussed in the previous subsection is given by $(-1)^J=(-1)^{ {\rm Ind}(\CD) }$.

Another way to describe anomalies is by the Atiyah-Patodi-Singer $\eta$ invariant,
which is defined by
\beq
\eta = \frac{1}{2} \left( \dim \ker \CD+ \sum_{\lambda \neq 0} {\rm sign}( \lambda)  \right),
\eeq
where the sum is taken over nonzero eigenvalues of $\CD$. However, on a closed manifold, 
the nonzero eigenvalues appear in pairs as $\pm \lambda$ as can be seen in \eqref{eq:antimatrix}, 
and hence their contributions are canceled out.
Therefore, we get
\beq
\exp( -2\pi i \eta)  = (-1)^{{\rm Ind}(\CD)},
\eeq
where we have used $(-1)^{\dim \ker \CD}=(-1)^{{\rm Ind}(\CD)}$.
The left-hand side is the anomaly formula of \cite{Witten:2015aba}, while the right-hand side is the anomaly formula of \cite{Witten:1982fp}.

\section{Global anomaly: the traditional method}
\label{sec:trad}

In this section, we compute the traditional global anomaly of the 8d
adjoint fermion using the methods of Elitzur and Nair
\cite{Elitzur:1984kr}, elaborating on an observation by Witten
\cite{Witten:1983tw}.

\subsection{A brief review of the strategy of Elitzur and Nair}

We will now briefly review the approach in
\cite{Witten:1983tw,Elitzur:1984kr}, to make the paper more
self-contained. (See also \cite{Lundell:1987bp} for a very clear and
detailed exposition of the procedure.)

The basic idea is to relate the computation of the global anomaly of a
representation $R_G$ of $G$ (which we assume to be free of local
anomalies) to a local anomaly under gauge transformations of a group
$F \supset G$. We choose $F$ so that there is some representation
$R_F$ of $F$ such that $R'_G$, appearing in the decomposition
$R_F=R_G\oplus R'_G$ under the embedding $G\subset F$, has known
anomalies. The simplest case is that $R'_G$ is a sum of copies of the
trivial one-dimensional representation, which clearly has no local or global
anomaly. Our aim is to relate the global anomaly of $R_G$ under $G$
transformations to the local anomaly of $R_F$ under $F$
transformations.

We do so as follows. The local anomaly under $f\in F$ (which means that $f$ is a gauge transformation in $F$, by abusing the notation) 
is the variation
of the phase of the fermionic path integral $Z_\psi$ on a manifold
$\CM_d$, which for simplicity we take to be $S^d$:
\begin{equation}
  Z_\psi[A] = e^{i\CA}Z_\psi[A^f]
\end{equation}
where $A$ is the connection on $\CM_d$, $A^f=fAf^{-1} + f df^{-1}$
its gauge transform, and we denote by $\CA$ the anomalous phase, which
we aim to determine. This can be done using descent, as usual, with
the result that
\begin{equation}
  \label{eq:A-descent}
  \CA = \int_{B_{d+1}} \!\!\! \mathrm{CS}(A^f) - \mathrm{CS}(A)
\end{equation}
where $\mathrm{CS}(A)$ is the Chern-Simons functional on $d+1$
dimensions for $A$, and $B_{d+1}$ is some $d+1$-dimensional manifold
with boundary $\CM_d=S^d$, which we take to be a ball. We have chosen
some arbitrary extension of $f$ to the interior of $B_{d+1}$. For
convenience we will introduce
\begin{equation}
\gamma(f,A)\equiv\mathrm{CS}(A^f)-\mathrm{CS}(A) \, .
\end{equation}

Consider now the specific case $f\in G$ when restricted to the
boundary of $B_{d+1}$, while allowing $f\notin G$ in the interior. We
denote this boundary value of $f$ by $g$. We also restrict $A$ to
belong to $G$. The anomalous phase of $Z_\psi$ under $f$ (viewed as an
element of $G$), i.e. the global anomaly that we are after, can then
be computed by~\eqref{eq:A-descent}.

Using elementary properties of the Chern-Simons functional, one can
verify that $\mathrm{CS}(A^g)=\mathrm{CS}(A)$ for $g\in G$, since we
are assuming that $G$ is free of local anomalies. This implies that
$\gamma(f,A)$ depends on $F$ only up to equivalence under
multiplication by elements of $G$, or equivalently it depends on $F/G$
only. Since we are setting $f=g\in G$ at the boundary, we can collapse
the boundary to a point for the purposes of
computing~\eqref{eq:A-descent}, and write
\begin{equation}
  \CA = \int_{S^{d+1}} \!\!\! \gamma(f,A)\, ,
\end{equation}
with the understanding that $f$ is valued in $F/G$ here. One can
easily check that this expression gives a homomorphism from
$\pi_{d+1}(F/G)$ to $\BR$. The normalization can be determined as
follows. Consider the long exact sequence in homotopy
\begin{equation}
  \label{eq:homotopy-LES}
  \ldots \to \pi_{d+1}(G) \to \pi_{d+1}(F) \to \pi_{d+1}(F/G) \to \pi_d(G) \to \ldots
\end{equation}
associated to the short exact sequence
$0\to G \to F \to F/G\to 0$. If we know the maps in the exact
sequence, and we know the normalization of $\CA$ for some of the
generators, we can work out (by linearity) the normalization for the
rest of the generators.

\subsection{Exercise in four dimensions}
Let us illustrate all this discussion with a simple example
\cite{Witten:1982fp,Witten:1983tw,Elitzur:1984kr}, where we choose
$d=4$, $G=\SU(2)$, $F=\SU(3)$, $R_F={\bf 3}$ and thus $R_G={\bf 2}$,
$R_G'={\bf 1}$. We have that $R_G\oplus R_G'$ is pseudo-real, so there
is no local anomaly for transformations in $G$. We want to determine
the global anomaly. The relevant homotopy groups are
$\pi_5(\SU(3))=\bZ$, $\pi_4(\SU(3))=0$, $\pi_4(\SU(2))=\bZ_2$, and
$\pi_5(\SU(3)/\SU(2))=\pi_5(S^5)=\bZ$.\footnote{The coset space
  $\SU(n)/\SU(n-1)$ is topologically $S^{2n-1}$.} We then read
from~\eqref{eq:homotopy-LES} that the following sequence is
exact\footnote{By ``0'' in homotopy exact sequences we mean the
  trivial group of one element. All the groups involved in our
  computations are abelian, which justifies the notation.}
\begin{equation}
  \ldots \to \underbrace{\pi_5(\SU(3))}_{\bZ} \xrightarrow{\,\alpha\,}
  \underbrace{\pi_5\biggl(\frac{\SU(3)}{\SU(2)}\biggr)}_{\bZ} \xrightarrow{\,\beta\,}
  \underbrace{\pi_4(\SU(2))}_{\bZ_2} \to 0\, .
\end{equation}
By exactness, it must be that $\alpha$ is multiplication by 2, and
$\beta$ is reduction modulo 2.

Consider now some choice of extension $\bar{\mathfrak{g}}$ of
$\mathfrak{g}$, the non-trivial generator of $\pi_4(\SU(2))$, into the
bulk of $B_5$. As before, we allow for the extension to be given by
arbitrary elements of $\SU(3)$ in the bulk. When projected down to
$\SU(3)/\SU(2)$, and after compactifying $B_5\to S^5$,
$\bar{\mathfrak{g}}$ gives us an odd multiple of the fundamental
generator $\mathfrak{q}$ of $\pi_5(\SU(3)/\SU(2))$.

Denote the fundamental generator of $\pi_5(\SU(3))$ by
$\mathfrak{f}$. For the case of interest, where we have a ${\bf 3}$ of
$\SU(3)$,
\begin{equation}
  \CA(\mathfrak{f}) = \int_{S^5} \gamma(\mathfrak{f})
\end{equation}
is simply the winding number,\footnote{Here we are abusing notation
  somewhat: $\CA$ is not the anomalous phase for the $\SU(3)$ theory,
  which would be given instead by the integral of $\gamma$ over a
  disk, not a sphere.} so $\CA(\mathfrak{f})=2\pi$. Since
$\alpha(\mathfrak{f})=2\mathfrak{q}$, and $\CA$ induces a
homomorphism, we learn that
$\CA(\mathfrak{q})=\CA(\bar{\mathfrak{g}})=\pi$, and there is a global
anomaly for the $\SU(2)$ transformation $\mathfrak{g}$, as first found
in \cite{Witten:1982fp}. More generally, due to the relation of
$\gamma$ to the $\SU(3)$ anomaly polynomial in six dimensions, we have
that $\CA(g)=2\pi I_3(R_F)$, where we define for $\SU(N)$, $N>2$
\begin{equation}
  \Tr_{R_F}(F^k) = I_k(R_F)\Tr_{\fund}(F^k)\, .
\end{equation}
with $d=2k-2$. We will use below the familiar fact that $I_k(R_F)=0$
if $R_F$ is real or pseudo-real and $k\in 2\BZ+1$.

\subsection{Computations in eight dimensions}

Let us now move on to the actual cases of interest: \Nequals1 theories in
eight dimensions. We have a fermion in the adjoint representation,
which has no local anomaly, but can potentially have a global anomaly
of the type just described whenever $\pi_8(G)\neq 0$. This is the case
for $G\in\{\SU(2),\SU(3),\SU(4),\SO(7)\ldots \SO(10),\SO(N),G_2,F_4\}$,
where we take $N>10$. See the list of homotopy groups in the appendix~\ref{sec:list}.
We have separated the $\SO(7)$ to $\SO(10)$ cases
from the rest since from $\SO(11)$ on the homotopy groups relevant for
our computation become stable.

\paragraph{The group $\SU(4)$:}

Let us consider first the $\SU(4)$ case, which we will embed in
$\SU(5)$. Using the fact that $\SU(5)/\SU(4)\simeq S^9$, and the known
homotopy groups of spheres and $\SU(n)$, we find that the following
portion of the homotopy long exact sequence is exact:
\begin{equation}
  \ldots \to \underbrace{\pi_9(\SU(5))}_{\bZ} \xrightarrow{\,\alpha\,}
  \underbrace{\pi_9\biggl(\frac{\SU(5)}{\SU(4)}\biggr)}_{\bZ} \xrightarrow{\,\beta\,}
  \underbrace{\pi_8(\SU(4))}_{\bZ_{24}} \to 0\, .
\end{equation}
As in the $\SU(2)$ case in four dimensions, we conclude that $\alpha$
is multiplication by 24, while $\beta$ is reduction modulo 24. The
adjoint of $\SU(5)$ is free of local anomalies in eight dimensions, so
$\gamma(f)=0$ in $\SU(5)$. The adjoint of $\SU(5)$ decomposes as
$\mathrm{Adj}_{\SU(4)}\oplus {\bf 4} \oplus \bar {\bf 4}\oplus {\bf
  1}$. Since ${\bf 4} \oplus \bar {\bf 4}\oplus {\bf 1}$ has no global
anomaly, we learn that the adjoint of $\SU(4)$ has no global anomaly
either.

\paragraph{The group $\SU(3)$:}

Using the embedding into $\SU(4)$, the relevant portion of the long
exact sequence in homotopy in this case is
\begin{equation}
  \begin{gathered}
    \begin{tikzpicture}[descr/.style={fill=white,inner sep=1.5pt}]
      \matrix (m) [
      matrix of math nodes,
      row sep=1em,
      column sep=2.5em,
      text height=1.5ex, text depth=0.25ex
      ]
      { \ldots & \pi_9(\SU(3)) & \pi_9(\SU(4)) & \pi_9(S^7) & \\
        & \pi_8(\SU(3)) & \pi_8(\SU(4)) & \pi_8(S^7) & 0 \\
      };

      \path[overlay,->, font=\scriptsize,>=latex]
      (m-1-1) edge (m-1-2)
      (m-1-2) edge (m-1-3)
      (m-1-3) edge (m-1-4)
      (m-1-4) edge[out=355,in=175] node[fill=white]{$\phi$} (m-2-2)
      (m-2-2) edge (m-2-3)
      (m-2-3) edge (m-2-4)
      (m-2-4) edge (m-2-5);
    \end{tikzpicture}
  \end{gathered}
\end{equation}
where we used $\SU(4)/\SU(3)\simeq S^7$. We have that
$\pi_9(\SU(3))=\bZ_3$, while $\pi_9(\SU(4))=\pi_9(S^7)=\bZ_2$. Exactness
of the sequence then implies that $\phi$ vanishes, and thus we end up
with the short exact sequence
\begin{equation}
  0 \to \underbrace{\pi_8(\SU(3))}_{\bZ_{12}} \xrightarrow{\,\alpha\,}
  \underbrace{\pi_8(\SU(4))}_{\bZ_{24}} \xrightarrow{\,\beta\,}
  \underbrace{\pi_8(S^7)}_{\bZ_{2}} \to 0\, .
\end{equation}
This implies that $\alpha$ is multiplication by 2, and $\beta$ is
reduction by 12. In particular the non-trivial generator of $\SU(3)$
maps to twice the non-trivial generator of $\SU(4)$, which we studied
in the section above. Decomposing the adjoint of $\SU(4)$ into $\SU(3)$
we conclude that the anomaly of the adjoint of $\SU(3)$ is given by
twice the anomaly of the adjoint of $\SU(4)$, which as explained above
vanishes.

\paragraph{Remark on $\SU(N)$:}
Physically, we can explain the above results for $\SU(3)$ and $\SU(4)$ as follows.
Let us consider $\SU(N)$ theory with an adjoint fermion. Let us also add a scalar in the adjoint representation. The scalar field does not contribute to the anomaly.
Now, suppose that this $\SU(N)$ theory is anomaly free. Then, by giving an appropriate expectation value to the adjoint scalar, 
we can break $\SU(N)$ to $\SU(N-1) \times \U(1)$. The fermion is now in the representation
$\mathrm{Adj}_{\SU(N-1)}\oplus {\bf N-1} \oplus \overline {\bf N-1}\oplus {\bf 1}$ of $\SU(N-1)$.
Since the original $\SU(N)$ theory was assumed to be anomaly free, the new $\SU(N-1)$ gauge group must also be anomaly free
under the RG flow from $\SU(N)$ to $\SU(N-1)$.
The fermions ${\bf N-1} \oplus \overline {\bf N-1}\oplus {\bf 1}$ do no contribute to the anomaly because we can add mass terms to them.
Thus we conclude that $\SU(N-1)$ with an adjoint fermion is anomaly free. For large enough $N$, $\pi_8(\SU(N))$ is zero.
Thus, we expect that $\SU(N)$ does not have anomaly associated to $\pi_8(\SU(N))$ for any small $N$, such as $N=3$ and $N=4$.

\paragraph{$\SO(N)$ for $N$ in the stable range:}

This case has been discussed in \cite{Zhang:1987pw}, also using the
general approach of \cite{Elitzur:1984kr}. We proceed by embedding
$\SO(N)$ into $\U(N)$. We choose $N$ large enough such that we are in
the stable range for all the homotopy groups entering our
computation. More concretely, we are assuming $N\geq 11$. The
relevant homotopy groups were computed in \cite{BottStable}. See also
\cite{Lundell1992} for a concise summary of the results used here, and
in following sections. From the long exact sequence in
homotopy~\eqref{eq:homotopy-LES} we have that
\begin{equation}
  0 \to \underbrace{\pi_{10}(\U/\SO)}_{\bZ_{2}} \xrightarrow{\,\alpha\,}
  \underbrace{\pi_9(\SO)}_{\bZ_{2}} \xrightarrow{\,\beta\,}
  \underbrace{\pi_9(\U)}_{\bZ} \xrightarrow{\,\gamma\,}
  \underbrace{\pi_9(\U/\SO)}_{\bZ} \xrightarrow{\,\delta\,}
  \underbrace{\pi_8(\SO)}_{\bZ_{2}} \to 0
\end{equation}
where we have omitted $N$ since it is irrelevant in the stable
range. Since necessarily $\beta=0$, we end up with the short exact
sequence in the right. We thus find that $\CA=\pi I_5(R_U)$. Since the
adjoint of $\SO(N)$ embeds as the antisymmetric of $\U(N)$, we find that
there is an anomaly whenever $I_5(\asymm)=N-16$ is odd, i.e. whenever
$N$ is odd.

\paragraph{The group $F_4$:}

We analyze $F_4$ by embedding into $E_6$. The quotient space $E_6/F_4$
(known as ``IV'' in Cartan's classification of symmetric spaces) has
well understood homotopy groups at low enough ranks \cite{Conlon}. In
particular, for $i\leq 15$ we have that
$\pi_i(E_6/F_4)=\pi_i(S^9)$. Proceeding as above, we end up with the
short exact sequence
\begin{equation}
  0 \to \underbrace{\pi_9(E_6)}_{\bZ} \xrightarrow{\,\alpha\,}
  \underbrace{\pi_9\biggl(\frac{E_6}{F_4}\biggr)}_{\bZ} \xrightarrow{\,\beta\,}
  \underbrace{\pi_8(F_4)}_{\bZ_{2}} \to 0\, .
\end{equation}
Consider first the branching ${\bf 27}\to {\bf 26}\oplus {\bf
  1}$. Since the ${\bf 27}$ of $E_6$ has a local anomaly, and
$\int\!\gamma=1$ for the generator of $\pi_9(E_6)$,\footnote{One way to see this may be to use $\SO(10) \subset E_6$ under which ${\bf 27} \to {\bf 10} \oplus {\bf 16}$.
We consider an $E_6$ bundle on $S^{10}$ such that $\SO(10) \subset E_6$ is identified with the tangent bundle of $S^{10}$.
From the index theorem, it is possible to show that $ {\bf 16}$ has one net zero mode while ${\bf 10}$ has no net zero mode, by using the fact that the 
Euler number of $S^{10}$ is 2.
} we conclude that
the ${\bf 26}$ representation of $F_4$ is anomalous. In turn, the
adjoint of $E_6$ decomposes as ${\bf 78}\to {\bf 26}\oplus {\bf 52}$,
where the ${\bf 52}$ is the adjoint representation of $F_4$. Since the
adjoint of $E_6$ gives rise to neither local or global anomalies, it
must be the case that the contribution of the ${\bf 26}$ cancels
against the contribution of the ${\bf 52}$. We thus learn that the
adjoint of $F_4$ has a global anomaly in eight dimensions.

\paragraph{The group $G_2$:}

We analyze $G_2$ by embedding into $F_4$, via the chain
$G_2\subset \SO(7) \subset \SO(9) \subset F_4$. The relevant branching
of representations are
\begin{align}
{\bf 52} & \to {\bf 14} \oplus {\bf 7}\oplus
\ldots\\
{\bf 26} & \to {\bf 7} \oplus \ldots
\end{align}
where the elided representations are either singlets or appear an even
number of times, which cannot give rise to a $\bZ_2$ valued global
anomaly. From the results in the previous section, we conclude that
the adjoint (${\bf 14}$) of $G_2$ is free of global anomalies in eight
dimensions.

\paragraph{Remaining cases: $\SU(2)$, $\SO(N)$ with $7 \leq N \leq 10$:}

These are somewhat more technical, but have been computed in
\cite{Zhang:1987pw} and papers cited therein. The result is that
$\SU(2)$ has no anomaly, and the $\SO(N)$ cases have anomaly as in the
stable case described above, i.e. whenever $N$ is odd.

\section{Global anomaly: a new method using instantons}
\label{sec:new}

In this section,
we discuss a way to find global anomalies of Weyl fermions in general representation $R_G$ of the gauge group $G$ 
in $d$ spacetime dimensions,
by introducing a codimension-4 gauge instanton.
Rather than attempting a completely general classification of anomalies,
here we just consider a specific setup to see the anomaly of $R_G$ and later discuss what kind of anomaly the specific setup is detecting.

\subsection{The basic idea}
\label{sec:instanton.method}

The basic idea is to consider a gauge theory soliton and see the
anomaly of the zero modes living on it.  For concreteness we focus our attention to the case where the gauge soliton is the familiar codimension-4 instanton.
However, the idea here is applicable to more general solitons.

Let us take a subgroup $\SU(2) \subset G$, and the maximal subgroup $H \subset G$
which commutes with $\SU(2)$.  Namely, we have
$[\SU(2) \times H]/\CC \subset G$, where $\CC$ is some subgroup of the
center of $\SU(2) \times H$.  An example is that
$[\SU(2) \times \SU(2)' \times \Spin(N-4)]/\BZ_2 \subset \Spin(N)$.
When the subgroup $\CC$ does not play any role, we will
often omit $\CC$ and loosely write $\SU(2) \times H \subset G$.

Suppose that the representation $R_G$ is decomposed under $\SU(2) \times H$ as
\beq
R_G  \to  \bigoplus_{n  \geq 1} (\mathbf{n}_{\SU(2)} \otimes R^n_H),
\eeq
where $\mathbf{n}_{\SU(2)}$ is the $n$-dimensional (i.e., spin $(n-1)/2$) irreducible representation of $\SU(2)$, 
and $R^n_H$ is some representation of $H$ which is not necessarily irreducible. 

If we consider an instanton of $\SU(2)$ with unit instanton charge, there are fermion zero modes living on it.
The representation $\mathbf{n}_{\SU(2)} $ produces $N_n=\frac{1}{6}(n^3-n)$ zero modes, where $N_n$ is twice the Dynkin index of $\mathbf{n}_{\SU(2)} $.
Then, we have a gauge group $H$ which is unbroken by the instanton, and the zero modes produce localized Weyl fermions in the representation
\beq
r_H :=  \bigoplus_{n  \geq 1} N_n  R^n_H . \label{eq:reducedrep}
\eeq

More concretely, let us suppose that the spacetime is $X= \BR^{d-4} \times S^4$,
and the instanton is put on $S^4$.
Then, at low energies, we get a $(d-4)$-dimensional gauge theory with the unbroken gauge group $H$ and Weyl fermions in $d-4$ dimensions in the representation $r_H$.
If this theory is anomalous, that means that the original theory is also anomalous.

If $R_G$ does not have a perturbative anomaly in $d$ dimensions, then
neither does $r_H$ in $d-4$ dimensions.  However, there can be global
anomalies as we will see explicitly later.

\subsection{Relation to traditional anomaly}

Now we study the relation of our anomaly associated to $r_H$ in $d-4$
dimensions and the traditional anomaly reviewed in section~\ref{sec:review}.  In the
traditional anomaly, the important point is that the gauge
configuration approaches to a pure gauge at the infinity of $\BR^{d+1}$.
However, we have considered the $S^4$ compactification in our discussion above,
so the relation of our anomaly to the traditional one is not obvious.

In a little more detail, the traditional anomaly can be detected by the mod-2 index (or more generally the $\eta$ invariant) on $S^{d+1}$ as reviewed in Sec.~\ref{sec:review},
where $S^{d+1}$ is the one point compactification of $\BR^{d+1}$.
On the other hand, the anomaly discussed above is detected by the mod-2 index (or the $\eta$ invariant) in $S^4 \times S^{d-3}$.
We put an $\SU(2)$ instanton on $S^4$, and also put a nontrivial gauge configuration of the gauge field of $H$ associated to $\pi_{d-4}(H)$ on $S^{d-3}$.

To connect the two anomalies, we need to do a nontrivial manipulation which we
now explain.
The main point is that if the condition 
\beq
\pi_{d-4}(G)=0. \label{eq:cond1}
\eeq 
is satisfied, then we can define a homomorphism \begin{equation}
\rho: \pi_{d-4}(H) \to \pi_d(G).
\end{equation}

Mathematically, this homomorphism is described as follows: 
On $S^{d+1}$, consider an embedding $S^{d-3} \subset S^{d+1}$ and take a tubular neighborhood $B^4\times S^{d-3} \subset S^{d+1}$ of $S^{d-3}$. 
The standard instanton bundle on $B^4$ and an $H$-bundle on $S^{d-3}$ 
defines a $G$-bundle on $B^4\times S^{d-3}$. The instanton bundle on $B^4$ is assumed to be pure trivial (i.e., pure gauge) on $\partial B^4=S^3$.
Since we assumed that $\pi_{d-4}(G)$ is trivial,
this $G$-bundle is trivial on the boundary $\partial(B^4\times S^{d-3})=S^3\times S^{d-3}$ because any element of $\pi_{d-4}(H)$ becomes zero in $\pi_{d-4}(G)$.
Therefore the bundle can be extended to the whole of $S^{d+1}$.
This defines an element in $\pi_d(G)$.

More physically, we can state the construction in the following way.
Our anomaly discussed above can be seen by considering a manifold $Y=S^{d-3} \times S^4$ as explained above.
Let us decompactify it to $\BR^{d-3} \times \BR^{4}$.
We put an instanton of $\SU(2)$ on $\BR^4$ which is localized near the origin, and also take a gauge configuration $B$ of the gauge group $H$ on $\BR^{d-3}$
such that it approaches to a pure gauge $B \to h^{-1}dh$ at the infinity of $\BR^{d-3}$, where $h$ corresponds to a nontrivial element of $\pi_{d-4}(H)$.
The relevant zero modes are localized near the intersection of these two configurations.

By taking the instanton size to be very small, we
get a codimension-4 object which we call ``instanton brane".\footnote{
This is motivated by the fact that small instantons on 7-branes give
D3-branes in string and F theory. However, our discussion does not
rely on string theory at all.}  On
$\BR^{d+1}=\BR^{d-3} \times \BR^{4}$, the instanton brane is localized
near $\BR^{d-3} \times \{0\}$, while the gauge field $B$ of the gauge
subgroup $H$ is localized near $ \{0\} \times \BR^4$.  Away from
$ \{0\} \times \BR^4$, the $B$ is trivial up to gauge transformations.

Now, suppose that the original gauge group $G$ satisfies the condition \eqref{eq:cond1}.
In this case, the $h \in \pi_{d-4}(H)$ becomes trivial in $\pi_{d-4}(G)$ and hence
we can almost deform the gauge field $B$ to the trivial configuration $B = 0$ as a gauge field of $G$.
However, the important point is that the deformation $B \to 0$ is possible only away from the instanton brane.
Near the instanton brane it is not guaranteed that we can make $B \to 0$
because the $\SU(2)$ instanton configuration may obstruct such deformation.
In this way, we get a gauge field configuration on $\BR^{d+1}$ which is localized near the submanifold $Z=\BR^{d-3} \times \{0\}$ 
and is trivial away from $Z$ up to gauge transformations. The gauge field $B$ is now localized near $\{ 0 \} \in \BR^{d+1}$.
This situation may be described as ``an instanton brane with a soliton inside it associated to $\pi_{d-4}(H)$".
The soliton inside the instanton brane supports the fermion zero modes relevant to our anomaly.

In the above argument, the world volume $Z$ of the instanton brane is extending to infinity.
However, the gauge field $B$ at infinity is of the form $h^{-1} dh$, and hence in the coordinate patch $\BR^{d-3} \setminus \{0\}$
we can make a gauge transformation such that $B=0$ near infinity. Then, the total gauge field configuration near infinity of $Z$
is simply that of the instanton brane without $B$.  Then we can compactify the world volume of the instanton brane 
to e.g., $Z=S^{d-3} \times \{0\} \subset \BR^{d-2} \times \BR^3$. 

In summary, we have obtained a gauge configuration which is localized on a compact submanifold $Z \subset \BR^{d+1}$.
Because $Z$ is compact, the gauge field must approach to a pure gauge at the infinity of $\BR^{d+1}$.
Therefore, this configuration is characterized topologically by an element of $\pi_d(G)$.
Namely, corresponding to each $h \in \pi_{d-4}(H)$, we get an element $\rho(h) \in \pi_d(G) $ if the condition \eqref{eq:cond1} is satisfied.
It may also be checked by a topological argument that this map 
\beq
\rho: \pi_{d-4}(H) \ni h \mapsto \rho(h) \in \pi_d(G)
\eeq
is a homomorphism from $\pi_{d-4}(H) $ to $\pi_d(G)$.
The map $\rho$ gives the relation between our anomaly and the traditional anomaly when the condition \eqref{eq:cond1} is satisfied.

If $r_H$ has a global anomaly under $h \in \pi_{d-4}(H)$, the instanton brane with the corresponding soliton
has odd number of fermion zero modes. This means that the original representation $R_G$ of the gauge group $G$ has an anomaly under $\rho(h) \in \pi_d(G)$
and hence the theory suffers from the traditional global anomaly.

How about the inverse direction? If the gauge configuration related to $\rho(h)$ does not give odd number of fermion zero modes for every $h$,
can we conclude that there is no global anomaly associated to $\pi_d(G)$?
At the level of the above argument, it is not possible because we have not yet shown that the map $\rho: \pi_{d-4}(H) \to \pi_d(G)$
is surjective. However, the surjectivity may be shown case by case. 
In practice, we encounter this problem in the case of the $G_2$ gauge group in $d=8$.
In this case, $\pi_8(G_2) = \BZ_2$. Now, the point is that $\rho$ is defined purely topologically, so it is independent of the representation $R_G$.
Thus, if we can find some representation $R_G$ which has the anomaly under $\rho(h)$, then that $\rho(h)$ must be a nontrivial element of  $\pi_8(G_2) = \BZ_2$.
This establishes the surjectivity of $\rho: \pi_{d-4}(H) \to \pi_d(G)$ in this particular case. 
We will show that the 7-dimensional representation of $G_2$ is really anomalous under $\rho(h)$.

For $d=8$, the only class of simple Lie groups for which \eqref{eq:cond1} is not satisfied is $\Sp(N)$.
For this case, our anomaly detected by $\pi_{d-4}(H)$ is actually different from the traditional anomaly associated to $\pi_d(G)$.
Indeed, we will see that 8d $G=\Sp(N)$ Super-Yang-Mills for $N \geq 2$ has the global anomaly related to $\pi_{4}(H)$, even though we have $\pi_8(\Sp(N))=0$ .
We will also argue that the anomaly of $\Sp(N)$ may be canceled by a TQFT, while
the anomaly associated to $\pi_d(G)$ cannot  be canceled by a TQFT.

\subsection{Exercise in four dimensions}

Let us see again Witten's original anomaly for $d=4$ and $G=\Sp(N)$ by using the argument in Sec.~\ref{sec:instanton.method}.
We present the argument for $G=\Sp(1)=\SU(2)$, but the generalization to $\Sp(N)$ is obvious. 

The maximal subgroup $H \subset \SU(2)$ which commutes with $\SU(2)$ is given by its center $H=\BZ_2$.
The condition \eqref{eq:cond1} is satisfied because $\pi_0(\SU(2))=0$.
Also, there is a possibility of an anomaly because $H=\BZ_2$ is discrete and $\pi_0(H)$ has two elements, trivial one and nontrivial one.

Let us consider the representation of $G=\SU(2)$,
\beq
R_G = \bigoplus_n c_n \mathbf{n}_{\SU(2)},
\eeq
where $\mathbf{n}_{\SU(2)}$ is the $n$ dimensional representation of $\SU(2)$ and $c_n$ are some non-negative integers.
Then the $r_H$ defined by \eqref{eq:reducedrep} is given by  
\beq
r_H = \left(  \sum_{n=\text{even}} c_n N_n \right) \bm{\epsilon}_{\BZ_2}+ \left(\sum_{n=\text{odd}} c_n N_n \right) \mathbf{1}_{\BZ_2}.
\eeq
where $\mathbf{1}_{\BZ_2}$ and $\bm{\epsilon}_{\BZ_2}$ are the trivial and nontrivial representations of $\BZ_2$, respectively.

Then we get a $d-4=0$ dimensional theory with $ \sum_{n=\text{even}} c_n N_n$ zero modes which transform nontrivially under the gauge subgroup $H=\BZ_2$.
There is anomaly of the path integral if $ \sum_{n=\text{even}} c_n N_n$ is odd. This can be seen, for example,
from the fact that the path integral in $d-4=0$ is an ordinary integral, and the integral measure is changed by $(-1)^{\sum_{n=\text{even}} c_n N_n}$
under the $H=\BZ_2$ gauge transformation. Therefore, the gauge invariance is violated if $\sum_{n=\text{even}} c_n N_n$ is odd.

For example, we have $N_2=1$, $N_4=10$, $N_6=35$ and so on. The fact that $\mathbf{4}_{\SU(2)}$ does not have global anomaly (because $N_4=\text{even}$) will be used later.

As discussed in general above, there is a homomorphism $\rho: \pi_0(\BZ_2) = \BZ_2 \to \pi_4(\SU(2))$. 
This map is constructed by considering a small pointlike $\SU(2)$ instanton whose world line sweeps $S^1$, and then including a nontrivial $H=\BZ_2$ holonomy
around this $S^1$. This construction was essentially discussed in Fig.~1 of \cite{Witten:1983tx} in a slightly different context.

\subsection{Computations in eight dimensions}
We study global anomalies of 8d Super-Yang-Mills (SYM) which automatically have maximal supersymmetry. 
The matter content can be easily seen from dimensional reduction of ten-dimensional (10d) SYM, which contains the gauge field and a Majorana-Weyl gaugino.
After reducing two dimensions, the Majorana-Weyl in 10d can be considered as Majorana (but not Weyl) in 8d which is equivalent to Weyl (but not Majorana) in 8d.
This is analogous to the fact that 4d gauginos can be considered as either Majorana or Weyl.
For our purposes, it is convenient to consider it as Weyl.

Before continuing, we note that the anomaly matching of 4d \Nequals2 theories was studied in \cite{Shimizu:2017kzs} on its Higgs branch which is assumed to have the form of the one-instanton moduli space of a Lie group $G$.
Such a 4d \Nequals2 theory would be realized on the worldvolume of the core of a BPST instanton in the 8d \Nequals1 supersymmetric $G$ gauge theory.
Therefore the analysis and the actual computation in that paper is very much related to those in this paper, and the results are consistent.

\paragraph{Simply laced groups:}
We do not discuss simply laced groups because of the following
reasons.  Simply laced groups other than $\SU(2)$ satisfies the condition \eqref{eq:cond1} given by $\pi_4(G)=0$,
so if there is an anomaly for some $h \in \pi_4(H)$, that anomaly comes
from $\rho(h) \in \pi_8(G) $ which cannot be canceled by TQFT, 
as already implicitly seen in the argument of \cite{Witten:1982fp,Witten:1985xe}
and as will be more explicitly explained in Sec.~\ref{sec:can't}.
However, we know that the SYM theories for all the simply laced groups appear
in F-theory.  Therefore, a priori it is expected that there is no
anomaly.  One can check it explicitly by considering some examples of
subgroups $\SU(2) \times H \subset G$.  For $G=\SU(2)$, we can only
have $H=\BZ_2$ and hence the anomaly associated to $\pi_4(H)$ is trivial.

\paragraph{The group $\SO(2N+1)$ with $N>2$:}
Let us consider $\SO$ (or more precisely $\Spin$) groups.
Let us take a subgroup
\beq
\SU(2) \times \SU(2)' \times  \SO(2N-3) ,
\eeq
and take $H =  \SU(2)' \oplus  \SO(2N-3)$. In this case, the adjoint representation ${\rm Adj}(\SO(2N+1))$ 
decomposes as 
\beq
\mathbf{2}_{\SU(2)} \otimes \mathbf{2}_{\SU(2)'} \otimes (\mathbf{2N-3})_{\SO(2N-3)} \oplus {\rm Adj}(\SU(2) \times \SU(2)' \times  \SO(2N-3))
\eeq
where $\mathbf{2}_{\SU(2)}$ and $\mathbf{2}_{\SU(2)'}$ are the $2$-dimensional (doublet) representations of $\SU(2)$ and $\SU(2)'$, respectively, 
and $(\mathbf{2N-3})_{\SO(2N-3)}$ is the $2N-3$ dimensional vector representation of ${\SO(2N-3)}$. 

One can compute $r_H$ defined in \eqref{eq:reducedrep} as
\beq
r_H = \mathbf{2}_{\SU(2)'} \otimes (\mathbf{2N-3})_{\SO(2N-3)} +(\text{$H$ singlets}).
\eeq
This representation contains an odd number (i.e., $2N-3$) of $\SU(2)'$ doublets, and hence suffers from the global anomaly in 4d.
Therefore, we conclude that $\SO(2N+1)$ has a global anomaly in 8d. This anomaly is associated to $\pi_8(\SO(2N+1))=\BZ_2$
as discussed above.

\paragraph{The group $G_2$:} The group $G_2$ contains a subgroup
\beq
\SU(2)_1 \times \SU(2)_2
\eeq
which can be seen from the affine Dynkin diagram. One can compute the decomposition of the adjoint representation and get the result
\beq
\mathbf{2}_{\SU(2)_1} \otimes \mathbf{4}_{\SU(2)_2} \oplus {\rm Adj}(\SU(2)_1 \oplus \SU(2)_2)
\eeq
Depending on which $\SU(2)$ to be taken as $H$,
we get either
\beq
r_H = \mathbf{4}_{\SU(2)_2} +(\text{$H$ singlets})
\eeq
for $H=\SU(2)_2$, or
\beq
r_H= 10 \cdot \mathbf{2}_{\SU(2)_1} +(\text{$H$ singlets})
\eeq
for $H=\SU(2)_1$.
In either case, there is no global anomaly in 4d.

One can also check that the homomorphism $\rho: \pi_4(H ) \to \pi_8(G)=\BZ_2$ is surjective. To see this,
take the 7-dimensional representation of $G_2$ which decomposes as
\beq
\mathbf{7}_{G_2} \to \mathbf{2}_{\SU(2)_1} \otimes \mathbf{2}_{\SU(2)_2} +\mathbf{3}_{\SU(2)_2}.
\eeq
This gives
\beq
r_H = \mathbf{2}_{\SU(2)'} +(\text{$H$ singlets})
\eeq
and hence the representation $\mathbf{7}_{G_2}$ suffers from the global anomaly associated to $\rho(h) \in \pi_8(G_2)$ for a nontrivial element $h \in \pi_4(\SU(2))$.
Because $\pi_8(G_2) = \BZ_2$, the $\rho$ must be surjective. Therefore, we conclude that the $G_2$ SYM theory does not have a traditional anomaly associated to
$\pi_8(G_2)$. It would be interesting to study whether this theory is completely anomaly free beyond the level of the traditional anomaly.

\paragraph{The group $F_4$:} The case $F_4$ can be treated easily by using the fact that it contains a subgroup $\SO(9) \subset F_4$
under which the adjoint representation decomposes as 
\beq
{\rm Adj}(F_4) \to {\rm Adj}(\SO(9)) \oplus \mathbf{2^4}_{\SO(9)},
\eeq
where $\mathbf{2^4}_{\SO(9)}$ is the $2^4$-dimensional spinor representation of $\SO(9)$.
We further take the subgroup $\SU(2) \times \SU(2)' \times \SO(5)$ as we did above for $\SO(2N+1)$.
One can check from the decomposition $\mathbf{2^4}_{\SO(9)} = \mathbf{2}_{\SU(2)}\otimes \mathbf{2^2}_{\SO(5)} + \mathbf{2}_{\SU(2)'}\otimes\mathbf{2^2}_{\SO(5)}$ that $ \mathbf{2^4}_{\SO(9)}$ does not contribute to the 4d anomaly of $\SU(2)'$.
Therefore, $F_4$ suffers from the global anomaly as in the case of $\SO(9)$.

\paragraph{The group $\Sp(N)$ with $N>1$:}
This class does not satisfy the condition \eqref{eq:cond1}, so the anomaly associated to
$\pi_4(H)$ is different from $\pi_8(G)$. Indeed, for $N>1$, we have $\pi_8(\Sp(N))=0$
and hence there is no traditional anomaly.

We take the subgroup
\beq
\Sp(1) \times \Sp(N-1) \subset \Sp(N),
\eeq
under which the adjoint representation decomposes as
\beq
{\rm Adj}(\Sp(N)) \to \mathbf{2}_{\Sp(1)} \otimes \mathbf{(2N-2)}_{\Sp(N-1)} \oplus {\rm Adj}(\Sp(1) \oplus \Sp(N-1)). 
\eeq
with obvious notations. Taking $H = \Sp(N-1)$ and introducing an instanton of $\SU(2)=\Sp(1)$, we get
\beq
r_H= \mathbf{(2N-2)}_{\Sp(N-1)}  +(\text{$H$ singlets}) .
\eeq
This is anomalous in 4d. Therefore, the SYM with the $\Sp(N)$ gauge group for $N>1$ suffers from some new
global anomaly which is different from the traditional one associated to $\pi_8(G)$.

However, the $\Sp(N)$ SYM is realized by an O7$^+$-plane in string
theory. Therefore, it must be possible to cancel the anomaly somehow.
Also, because of the tight constraints from supersymmetry, there is no
freedom to add local propagating degrees of freedom.  Therefore, the
anomaly must be canceled by coupling to a TQFT.

\section{Uncancellability of traditional global anomaly}
\label{sec:can't}
In view of the recent developments on anomalies, one can ask whether
the traditional anomaly associated to $\pi_d(G)$ can be canceled by
coupling to a TQFT. We here argue that this is not possible.  In this
section, the gauge field is treated as a background field because its path integral plays no role.
Essentially we follow the original arguments of \cite{Witten:1982fp,Witten:1985xe}, with the possible existence of a TQFT in mind.

Let $a$ be some gauge field configuration (possibly trivial) which has compact support on $\BR^d$. Let $g$ be a gauge transformation representing
a nontrivial element of $\pi_d(G)$ which also has compact support on $\BR^d$. Let $a^g=g^{-1} a g +g^{-1} dg$ be a gauge transform of $a$ by $g$, which
again has compact support. As reviewed in Sec.~\ref{sec:pid.anomaly}, the traditional anomaly can be seen by going from $a$ to $a^g$ by a path like $(1-f(t))a+f(t)a^g $
where $f(t \to -\infty)=0$ and $f(t \to +\infty)=1$.
The anomaly is represented by the change of the phase of the fermion partition function 
$\det P_L\slashed{D}[a(t)]$ under this continuous local deformation of the gauge field. 

The question is whether such an anomaly can also be produced by a theory without massless propagating degrees of freedom.
We denote the partition function of such a gapped theory as $Z_{\rm TQFT}[a]$.
This notation implies that the low energy limit is described by a TQFT.

The locality principle in quantum field theory suggests the following.
If a theory is gapped, the change of the partition function under any local continuous deformation of background fields can be 
captured by a local effective action in the low energy limit. 
This can be seen by the following argument. Under a small (i.e., topologically trivial, but not necessarily infinitesimal) deformation $\delta a$ of the field $a$,
the change of the partition function is given by  
\beq
\frac{Z_{\rm TQFT}[a + \delta a]}{Z_{\rm TQFT}[a]}=\vev{ \exp( i \int J^\mu \delta a_\mu)} \label{eq:smallchange}
\eeq
where $J^\mu$ is a local operator coupled to $\delta a$ (or more explicitly the current operator to which the gauge field is coupled).
There may be other terms of the form $(\delta a)^2 K + \cdots $ in the action, but the argument below is the same even if we include them.
By expanding  $\delta a$, it is reduced to the computation of correlation functions of $J^\mu$.
In a gapped theory, the correlation functions decays exponentially fast,
and in the limit of very large mass gap, they are given just by contact terms. 
This means that the result of \eqref{eq:smallchange} is given by a local polynomial (possibly with derivatives) of $\delta a$.
Notice that there is no room for TQFT to change this conclusion, because we are just considering topologically trivial deformation $\delta a$
which has compact support on $\BR^d$.

As a result of the above argument, the effective action defined by
\beq
S[a]=\log Z_{\rm TQFT}[a] - \log Z_{\rm TQFT}[0]
\eeq
is a local polynomial action of $a$ (neglecting irrelevant higher dimensional operators) for a topologically trivial $a$.\footnote{
We remark that there is a difference between a local action and a local polynomial action. To explain this point and also to illustrate the argument below \eqref{eq:smallchange},
let us consider a gapped theory in odd dimensions in which the $Z_{\rm TQFT}[a]$ is given by $e^{-2\pi i \eta}$, 
where $\eta$ is Atiyah-Patodi-Singer $\eta$ invariant of some Dirac operator coupled to $a$. See e.g., \cite{Witten:2015aba,Yonekura:2016wuc } for discussions on such a theory.
Under a local continuous deformation, the change of $\eta$ is captured by the change of Chern-Simons action.
Notice that $\eta$ itself is not represented by a polynomial of $a$. 
Only the difference $S[a]=\log Z_{\rm TQFT}[a]-\log Z_{\rm TQFT}[0]$ for topologically trivial $a$ can be represented by a Chern-Simons action which is a polynomial of $a$.
However, the $\eta$ invariant is local in some appropriate sense because it satisfies the gluing law~\cite{Dai:1994kq} (see \cite{Yonekura:2016wuc} for physics explanation). 
Our action $S[a]$ is a polynomial of $a$ as the argument below \eqref{eq:smallchange} clearly shows. 
We can enumerate such local polynomial actions.}
Moreover, the absence of perturbative anomaly implies that $S[a]$ is invariant under infinitesimal gauge transformations.

Remember that both $a$ and $a^g$ are topologically trivial on $\BR^d$ with compact support.
Namely, they are not just pure gauge, but are literally zero outside a compact region. 
Then the difference of the logarithm of the partition functions between $a$ and $a^g$ is given by
\beq
\log Z_{\rm TQFT}[a^g] - \log Z_{\rm TQFT}[a] = S[a^g] - S[a],
\eeq
However, in even dimensions, there is no such local polynomial action $S[a]$ which is invariant under infinitesimal gauge transformations,
and which produces the anomaly $S[a^g] - S[a]  \neq 0 $ for nontrivial element $g \in \pi_d(G)$.
Instead, we have $S[a^g] = S[a] $.
We conclude that the anomaly associated to $\pi_d(G)$ in $d=$~even dimensions cannot be produced by a gapped system without massless propagating degrees of freedom.
This excludes the possibility that the traditional anomaly can be canceled by a TQFT.

In odd dimensions, a Chern-Simons action can have
$S[a^g] - S[a] \neq 0 $. This is relevant for the parity anomaly if we
regard it as a kind of global anomaly.  However, the Chern-Simons
action is just a local action, and a TQFT
does not play any role. Therefore, the global anomaly which is not
cancelled by a Chern-Simons counterterm cannot be cancelled by a TQFT
in odd dimensions either.

\section{Topological Green-Schwarz mechanism}
\label{sec:topGS}
We have seen that the 8d $\Sp(N)$ SYM suffers from the new global anomaly which is different from the traditional one associated to $\pi_8(G)$. Since the 8d $\Sp(N)$ SYM can be realized in string theory, the anomaly must be canceled, and the cancellation must be carried out by coupling to a TQFT. In this section, we discuss how this kind of subtle anomaly might be sometimes canceled by a TQFT in a way analogous to the standard Green-Schwarz mechanism. 

In Sec.~\ref{sec:heuristics}, we first explain an analogue of the Green-Schwarz mechanism which uses topological degrees of freedom, using a TQFT which couples to ordinary homology cycles. 
Then, in Sec.~\ref{sec:3d}, we discuss an example where this method actually works in 3d: an $\SU(N)/\bZ_N$ gauge theory coupled to an adjoint.
But unfortunately this does not work in 8d $\Sp(N)$ theory, as we explain in Sec.~\ref{sec:K}.
We then discuss some of the expected properties of the 8d TQFT.
Most of the contents of Sec.~\ref{sec:heuristics} and Sec.~\ref{sec:3d} are not new,
but we put emphasis on some of the points relevant to the discussion of the 8d $\Sp(N)$ anomaly.

\subsection{Topological Green-Schwarz mechanism for cohomological classes}
\label{sec:heuristics}
Suppose $X$  and $Y$ are some characteristic classes in cohomology with $\BZ_k$ coefficient which are associated to gauge fields of group $G$. 
If we are given a manifold $M$ equipped with gauge fields, then we get 
elements of cohomology $X \in H^\bullet(M ; \BZ_k)$ and $Y \in H^\bullet(M ;\BZ_k)$.

Let $M$ be a $d$-dimensional manifold with $G$-bundle $P$. We also assume that $X \in H^p(M ;\BZ_k)$
and $Y \in H^{d-p+1}(M ;\BZ_k)$ for an integer $p$.
Then we want to consider a $(d-p)$-form $\BZ_k$ gauge theory coupled to $X$ and $Y$, which is roughly described by
\beq
S  =  ``~ 2\pi i \int_M \left( k b' \d a'  + b' X  +(-1)^{d-p+1} Y a' \right) ",
\eeq
where $b'$ and $a'$ are $(d-p)$-form and $(p-1)$-form fields,\footnote{More precisely, they must be treated by using differential cohomology theory, but we will be sloppy throughout the paper
and just pretend as if they were the usual differential forms.
This does not affect the final conclusion.}
and $\d$ is the exterior derivative. 
However, this action still does not make sense. This is because $X$ and $Y$ are elements of cohomology $H^\bullet(M ; \BZ_k)$, but, at the level of classical Lagrangian, $a'$ and $b'$
are differential forms which are not necessarily closed. Eventually this problem leads to the anomaly involving $X$ and $Y$ as discussed e.g.~in \cite{Kapustin:2014tfa}.

We give an argument which is slightly different from \cite{Kapustin:2014tfa}.
For simplicity, we assume that the manifold $M$ has no torsion in its cohomology so that 
$H^\bullet(M ; \BZ_k) = H^\bullet(M ; \BZ) \otimes \BZ_k$.

First, we change the variables from $a'$ and $b'$ to $a$ and $b$ such that
\beq
f:=\d a = \d a' +\frac{1}{k}X,~~~~~g:=\d b = \d b'+\frac{1}{k} Y.
\eeq
The fluxes $f= \d a$ and $g = \d b$ are gauge invariant. 
The effects of the background fields $X$ and $Y$ are now incorporated in quantization conditions of the fluxes $f$ and $g$ as
\beq
[f] \in \frac{1}{k}X +  H^p(M ;\BZ),~~~~~[g] \in \frac{1}{k}Y + H^{d-p+1}(M ;\BZ). \label{eq:quantization}
\eeq
where $[f]$ and $[g]$ are cohomology classes represented by $f$ and $g$. Namely, the fractional part of the fluxes of $f$ and $g$ are determined by $X$ and $Y$, respectively.
These shifts of the quantization conditions are analogous to the shift of 't~Hooft magnetic flux in the presence of background fields of 1-form center symmetry
in Yang-Mills theory~\cite{Kapustin:2014gua, Gaiotto:2014kfa}. In the present case, the relevant symmetries are higher form (or more explicitly $(d-p)$-form and $(p-1)$-form)
$\BZ_k$ symmetries.

To obtain a gauge invariant action, we follow the standard procedure. Let $N$ be a $d+1$ dimensional manifold whose boundary is our $d$-manifold $M$.
We assume that $X$ and $Y$ can also be extended to $N$.
In this situation, we take the action as
\beq
S_N = 2\pi i k \int_N   g f   .\label{eq:higherTQFT}
\eeq
This action is gauge invariant. However, it depends on how we extend $M$ and the cohomology classes $X$ and $Y$ on it to the manifold $N$. 
The standard way to see it is to consider another manifold $N'$ and define the action $S_{N'}$. The difference $S_N - S_{N '}$
is given by considering a closed manifold $L$ which is obtained by gluing $N$ and $N'$ along their common boundaries, and evaluating 
$
S_L= 2\pi i k \int_L   g f .
$
Now, because of the quantization conditions \eqref{eq:quantization}, this integral on the closed manifold $L$ is evaluated to be
\beq
S_L = \frac{2\pi i}{k} \int_L YX.
\eeq
This is the anomaly of this system. 

In terms of the topological defect operators, the $(p-1)$ form symmetry whose background field is $X$ and the $(d-p)$-form symmetry whose background is $Y$
have associated `volume operators' supported on $(d-p)$-dimensional submanifolds 
and on $(p-1)$-dimensional submanifolds, respectively.
This anomaly means that they have a braiding phase $\exp(2\pi i/k)$.

Now, suppose that we consider a gauge theory of a gauge group $G$ with some Weyl fermions.
If the gauge theory has a global anomaly given by $-\frac{2\pi i}{k} \int_L YX$ for some characteristic classes $Y$ and $X$,
then we can cancel the anomaly by adding the TQFT given by \eqref{eq:higherTQFT}. This is what we mean by topological Green-Schwarz mechanism.

There are also important topological constraints coming from the TQFT. The equations of motion of \eqref{eq:higherTQFT} are given by $f=0$ and $g=0$.
However, because of the quantization conditions \eqref{eq:quantization}, we have to impose the conditions that $X$ and $Y$ are trivial as elements of $H^\bullet(M; \BZ_k)$.
These constraints are analogous to the equation $dH=\tr RR - \tr FF$ in heterotic string theories, where $H$ is the field strength of the NS 2-form field,
$R$ is the Riemann curvature, and $F$ is the field strength of the heterotic gauge group. This equation requires that $\tr RR - \tr FF$ is trivial in de~Rham cohomology.

In our 8d $\Sp(N)$ SYM, the above topological constraints must work as follows, assuming the existence of some appropriate TQFT which cancels the anomaly of $\Sp(N)$.
In Sec.~\ref{sec:new}, we found the anomaly by compactifying the theory on $S^4$ and putting an instanton on it.
The TQFT must forbid such a configuration. Namely, the necessary condition on the TQFT is that it must give the constraint that the instanton number on $S^4$ is even.
So the TQFT gives the constraints on the sum over topological sectors~\cite{Seiberg:2010qd}.

\subsection{Topological Green-Schwarz mechanism in action: a 3d example}
\label{sec:3d}
Now let us discuss an example where the topological Green-Schwarz mechanism as described above is in action.
Consider the 3d Majorana fermion in the adjoint of $\mathfrak{su}(N)$, where we assume $N$ is even.
It has an anomaly captured by the 4d topological term 
\begin{equation}
2\pi i \cdot \frac{N}{2}\int c_2(F) \label{4dFF}
\end{equation} where $c_2(F) \propto \tr F \wedge F$ is the second Chern class in the normalization where it integrates to one for the $\SU(N)$ one-instanton. (A complex fundamental fermion needs $\pm1/2$ Chern-Simons. An adjoint Majorana fermion then needs level $\pm 1/2 \cdot  2N \cdot 1/2= N/2$.)
Now we try to gauge it with $\SU(N)/\bZ_{N}$. 
Let us impose time-reversal invariance so that we cannot add any Chern-Simons counterterms.
Then this is anomalous,  since a configuration of $\SU(N)/\bZ_{N}$ can have  instanton number $1/N$.
This is a version of the so-called parity anomaly, but we are imposing time-reversal invariance and hence the gauge symmetry is anomalous.

But the anomaly \eqref{4dFF} can also be written modulo $2\pi i$ as \begin{equation}
-2\pi i \cdot\frac{1}{4}\int w_2\wedge w_2 \label{eq:w2w2}
\end{equation}
where $w_2\in H^2(M_4,\bZ_{N})$ is the (generalized) Stiefel-Whitney class of the bundle.\footnote{More precisely, 
for $N \equiv 2 \mod 4$,
we need to use the Pontryagin square operation to make sense of the factor $1/4$, see e.g.~Sec.~6 of \cite{Aharony:2013hda} or \cite{Kapustin:2013qsa}.
If we assume that the manifold has no torsion in cohomology, we can lift $w_2$ to an element of integer cohomology and define $w_2 \wedge w_2$ there.
}
Therefore,  a 3d TQFT which couples to a background $\bZ_{N}$ one-form symmetry with this anomaly can cancel it.

Note that in terms of the line operator coupled to $w_2$, the anomaly \eqref{eq:w2w2} just means that the self braiding phase is $i = \exp(2\pi i/4)$. 
In other words, the topological spin of the operator is $1/4$, which is detected by ``twisting" the line operator~\cite{Witten:1988hf}.

If $N$ is a multiple of 4, the anomaly can be canceled by using the $\BZ_4$ theory with $d=3, p=2$ and taking $ X,Y$ to be the mod 4 reduction of $w_2$.
However, in the present case of the anomaly of the form $X^2$ (with $X=w_2$), there is more economical choice which is applicable to any even $N$. 
We use the $\U(1)_{2} \times \U(1)_{-1}$ Chern-Simons theory which is 
time-reversal invariant~\cite{Seiberg:2016rsg, Tachikawa:2016cha, Tachikawa:2016nmo}. It has a $\BZ_2$ 1-form symmetry which can be coupled to
the mod 2 reduction of $w_2$, and the coupling can be done as a shifted quantization condition of the $\U(1)_2$ gauge field as in \eqref{eq:quantization}.
The anomaly is computed exactly as in the previous subsection, and is given by \eqref{eq:w2w2}.

\subsection{The need for a subtler version in 8d}
\label{sec:K}

The idea above does not exactly work in our 8d $\Sp(N)$ theory if we insist on using ordinary (co)homology.
To see this, recall how we found the anomaly in Sec.~\ref{sec:new}.
As reviewed in Sec.~\ref{sec:review}, the anomaly is given by the number of zero modes modulo 2 in a 9-dimensional manifold $N$. To find an anomaly, in Sec.~\ref{sec:new}
we first considered the instanton brane $Z$ which is realized by codimension-4 instanton of $\Sp(N)$ embedded in $\SU(2) \subset \Sp(N)$.
This has codimension 4 and hence $\dim Z =5$.
There are localized zero modes on $Z$ in the fundamental representation of $\Sp(N-1) \subset \Sp(N)$. 
These modes can be regarded as fermions living on the brane $Z$ coupled to $\Sp(N-1)$.
Then, we further evaluate the mod 2 index of these fermions on $Z$.

The location of the instanton brane $Z$ at the homological level is determined by the Poincare dual of $q_1$ which is
the Pontryagin class or the instanton number of the $\Sp(N)$ bundle (which is equivalent to the second Chern class $c_2$ of the fundamental representation of $\Sp(N)$).
Thus, naively, the topological Green-Schwarz mechanism requires
\beq
\text{naively:}~~~~~X \sim q_1,~~~Y \sim \text{[mod 2 index in 5d]}.
\eeq
Then the anomaly is described as
\beq
\text{naively:}~~~~~\pi i \int  XY \sim \pi i \int_Z \text{[mod 2 index in 5d]}
\eeq
by using the Poincare duality $q_1 \leftrightarrow Z$.

However, we need more information than ordinary (co)homology to compute the mod 2 index.
In other words, there is no formula which gives the mod 2 index in terms of the cohomological characteristic classes of the gauge bundle (and metric).
This can be seen from the fact that we can have a nontrivial mod 2 index on $S^5$ as shown in the original paper of global anomaly~\cite{Witten:1982fp},
but there is no cohomological characteristic classes because the classifying space $B\Sp(N)$ of the $\Sp(N)$ group has $H^5(B\Sp(N))=0$.\footnote{ 
The cohomology $H^\bullet(B\Sp(N),\BZ)$ is freely generated as a ring by Pontryagin classes $1,q_1, q_2, \cdots, q_N$ where $q_i \in H^{4i}(B\Sp(N),\BZ)$. 
Cohomology with more general coefficients can be obtained by universal coefficients theorem from $H^\bullet(B\Sp(N),\BZ)$.
Also, there is no nontrivial gravitational characteristic classes on $S^5$ which can mix with $q_1$ to give a nontrivial value.}

Moreover, to specify the instanton brane $Z$, it is important that the fermions on $Z$ has a definite spin structure because the mod 2 index depends on it.
However, that information is missing in the homology class of $Z$ determined by the Poincare dual of $q_1$.
The spin structure is determined as follows.
The instanton has an anti-self-dual curvature, and in particular locks the $\SU(2)$ gauge bundle and the normal bundle to the locus of the instanton brane $Z$.
This means that the holonomy of the normal bundle is reduced from $\SO(4)$ to $\SU(2)$.
This reduction of the structure group of the normal bundle, combined with the spin structure on the total space time, gives the spin structure to the normal bundle, and hence to the tangent bundle.

The above considerations suggest that we need a concept of Poincar\'e duality which gives us a definite spin structure on the submanifold $Z$.
Indeed mathematicians have developed such a concept: KO-homology of Baum-Douglas, see e.g.~string theory articles which use them, \cite{Szabo:2007wp,Valentino:2008xd}.

The essential idea behind it is as follows. 
In the case of ordinary $\bZ$-coefficient homology $u\in H^i(M_d,\bZ)$ on a $d$-manifold $M_d$,
an orientation of $M_d$ defines the `volume form'  $[M_d]\in H_d(M_d,\bZ)$,
and then $\PD[u]\in H_{d-i}(M_d,\bZ)$.
Furthermore, for $v\in H^{j}(M_d,\bZ)$, $\int_{[M_d]} u v= \int_{\PD[u]} v$.

These concepts generalize to K and KO-(co)homologies. 
First, the KO-homology defines groups $KO_i(M_d)$ for a given positive integer $i$ and a $d$-dimensional manifold $M_d$, as the ordinary homology $H_i(M_d)$ does. There is also KO-cohomology groups, denoted by $KO^i(M_d)$.
A KO-orientation of $M_d$ is a spin structure and defines the KO-theoretic `volume form' $[M_d]\in KO_d(M_d)$.
Then, a class $u\in KO^i(M_d)$ has a Poincar\'e dual $\PD[u]\in KO_{d-i}(M_d)$,
such that $\int_{[M_d]} uv = \int_{\PD[u]} v$.
Here, the integration symbol in KO-theory means that we take the index of the appropriate Dirac operator.

In the present context, an $\Sp(N)$ bundle $P$ on a 9-manifold $M_9$ gives an element $[P]$ of the symplectic K-theory on $M_9$, which is canonically isomorphic to $KO^4(M_9)$ by Bott periodicity.
\footnote{There is also a symplectic version of K-(co)homology called KSp-(co)homology, and Bott periodicity relates $KSp$ and $KO$ by $KSp^i$=$KO^{i-4}$=$KO^{i+4}$.}
Its Poincare dual $\PD[P]\in KO_{5}(M_9)$ in the KO sense is exactly our instanton brane $Z$ equipped with the spin structure described above.

An important formal difference between the integration of KO theory and the integration in ordinary (co)homology is that in the latter, to have non-vanishing $\int_{[M_d]} u$, 
the $u$ needs to be in $H^d$.
This is no longer the case for the KO theory: in general, 
\begin{equation}
\int_{[M_d]} u \in KO^{i-d}(\mathrm{pt})\quad\text{for}\quad u\in KO^i(M_d)
\end{equation}
where $KO^{i}(\mathrm{pt})$ is a KO-cohomology on a single point ${\rm pt}$.
We have \begin{equation}
\begin{array}{c|ccccccccccc}
i & 0 & -1 & -2 & -3 & -4 & -5 & -6 & -7 \\
\hline
KO^i & \BZ & \BZ_2 & \BZ_2 & 0 & \bZ & 0 & 0& 0 
\end{array}
\end{equation}
and $KO^i =KO^{i+8}$.

For example, the mod-2 index of an Sp bundle $P$ over a spin 5-manifold $M_5$ can be understood as the integration in KO theory. 
Indeed, $\xi:=[P]\in KO^4(M_5)$, and therefore $\int_{M_5} \xi  \in KO^{-1}(\mathrm{pt})=\bZ_2$. 
This can be evaluated in two steps: the Poincar\'e dual of $\xi$ is an element $\PD[\xi]\in KO_1(M_5)$, which is a one-cycle with a spin structure.
Then we have \begin{equation}
\int_{M_5} \xi = \int_{\PD[\xi]} 1 \in KO^{-1}(\mathrm{pt})=\bZ_2.
\end{equation} 
Note that the mod-2 index of a circle with a spin structure is $0$ if the spin structure is NS (anti-periodic) and $1$ if it is R (periodic).
The class $\xi$ also gives instanton numbers when integrated over 4-manifolds.
Therefore, it plays the role of both $q_1$ and [mod 2 index in 5d]. So, very roughly speaking,
we may identify $\xi \sim X+Y$.

Let us come back to the question of the 8d Sp theory.
Given an Sp bundle $P$ on a 8d spin manifold $M$, 
we consider a fermion in the adjoint representation.
There is no perturbative anomaly and no global anomaly in the traditional sense associated to $\pi_8(\Sp)$.
There is however an anomaly, whose phase is characterized by \begin{equation}
\int_N \Sym^2\xi \in \bZ_2\label{XXX}
\end{equation} where $N$ is the 9d manifold,
$\xi$ is the class in $KSp^0(N)=KO^4(N)$ associated to the vector bundle in the fundamental representation of an Sp bundle $P$ over $N$,
and $\Sym^2: KO^4 \to KO^8 = KO^0$ is the symmetric power sending the fundamental representation of Sp to the adjoint representation.
Very roughly, we may think of it as $\Sym^2\xi \sim \frac{1}{2} \xi^2 \sim XY$.

So, we need a 8d TQFT such that 
\begin{itemize}
\item it can couple to elements $\xi\in KO^4(M)$, and
\item it has an anomaly characterized by \eqref{XXX}. 
\end{itemize} 
Note that if it can couple to an ordinary cohomology $\xi\in H^4(M,\bZ_k)$ instead,
we would have said that this TQFT has a $\bZ_k$ 3-form symmetry.
Since it can couple to an element in $KO^4(M)$, it has some generalized notion of symmetry.

The anomaly \eqref{XXX} suggests that the required TQFT is a KO-theoretic version of abelian Chern-Simons theory
which is somewhat analogous to the theory $\U(1)_2 \times \U(1)_{-1}$ mentioned at the end of Sec.~\ref{sec:3d}.
The authors hope to come back to this problem in the future.

\section*{Acknowledgments}
I.G.-E. thanks Miguel Montero and Diego Regalado for discussions. KO gratefully acknowledges support from the Institute for Advanced Study.
YT is partially supported in part byJSPS KAKENHI Grant-in-Aid (Wakate-A), No.17H04837 
and JSPS KAKENHI Grant-in-Aid (Kiban-S), No.16H06335, and 
also supported in part by WPI Initiative, MEXT, Japan at IPMU, the University of Tokyo.
The work of KY is supported in part by the WPI Research Center Initiative (MEXT, Japan), and also supported by JSPS KAKENHI Grant-in-Aid (17K14265).

\appendix

\section{List of homotopy groups}\label{sec:list}
Table~\ref{table:homotopy} lists low-degree homotopy groups of compact Lie groups.
The table itself can be found in ~the Appendix of \cite{EDM}.
For computations, see \cite{MT}.
We only consider simply connected groups $\pi_1(G)=0$,
while all compact simple Lie groups have $\pi_2(G)=0$ and $\pi_3(G)=\BZ$. 
$\pi_4(G)$ is trivial except for $G=\Sp(N)$, in which case  $\pi_4(G)=\bZ_2$.
$\pi_4(G)$ also has a derivation uniform to all $G$ in terms of the root system, see \cite{BS}.\footnote{The authors thank the discussions at \url{https://mathoverflow.net/questions/259487/}.}

For the infinite series $\SU(N)$, the homotopy groups $\pi_d(\SU(N))$ become stable for $N > d/2$.
For the infinite series $\Spin(N)$, the homotopy groups $\pi_d(\Spin(N))$  become stable for $N>d+1$.
For the infinite series $\Sp(N)$, the homotopy groups $\pi_d(\Sp(N))$  become stable for $N>(d-2)/4$.

\begin{table}[ht]
\[
\begin{array}{c|cccccccccccccc}
G~\setminus~d&4&5&6&7&8&9&10&11 \\
\hline
\hline
\Sp(1) & \BZ_2 &\BZ_2 & \BZ_{12} &\BZ_{2} &\BZ_{2} &\BZ_{3} & \bZ_{15} &\bZ_2\\
\Sp(2) &  \BZ_2 &\BZ_2 & 0 &\BZ_{} &0 &0 & \bZ_{120} &\bZ_2\\
\Sp(N \geq 3) &  \BZ_2 &\BZ_2 & 0 &\BZ_{} &0 &0 & 0 & \bZ\\
\hline
\SU(3) & 0 &\BZ &\BZ_{6} &0 &\BZ_{12} &\BZ_{3} & \bZ_{30} & \bZ_4 \\
\SU(4) &0 &\BZ &0 &\BZ &\BZ_{24} &\BZ_{2} &   \bZ_{120}\times \bZ_2 & \bZ_4 \\
\SU(5) &0 &\BZ & 0 &\BZ &0 &\BZ &  \bZ_{120}& 0 \\
\SU(N \geq 6) &0 &\BZ & 0 &\BZ &0 &\BZ &  0 & \bZ  \\
\hline
\Spin(7) &0 &0 &0 &\BZ_{} &(\BZ_{2})^2 &(\BZ_{2})^2 & \bZ_8 & \bZ\times \bZ_2 \\
\Spin(8) &0 &0 &0 &\BZ_{}^2 &(\BZ_{2})^3 &(\BZ_{2})^3 &\bZ_{24}\times \bZ_8 & \bZ\times \bZ_2    \\
\Spin(9) &0 &0 &  0 &\BZ_{} &(\BZ_{2})^2 &(\BZ_{2})^2 &   \bZ_8 & \bZ\times \bZ_2\\
\Spin(10) &0 &0 & 0 &\BZ_{} &\BZ_{2} &\BZ \times \BZ_{2} &  \bZ_4 & \bZ \\
\Spin(11) &0 &0 &0 &\BZ_{} &\BZ_{2} &\BZ_{2} &   \bZ_2 & \bZ\\
\Spin(12) &0 &0 &0 &\BZ_{} &\BZ_{2} &\BZ_{2} &   0 & \bZ\times \bZ \\
\Spin(N \geq 13) &0 &0 &0 &\BZ_{} &\BZ_{2} &\BZ_{2} &   0 & \bZ\\
\hline
G_2 &0 &0 & \BZ_{3} &0 &\BZ_{2} &\BZ_{6} &   0 & \bZ\times \bZ_2\\
F_4 &0 &0 &  0 &0 &\BZ_{2} &\BZ_{2} &0 & \bZ\times \bZ_2 \\
E_6 &0 &0 &  0 &0 &0 &\BZ & 0&\bZ \\
E_7 &0 &0 &  0 &0 &0 &0 &  0 & \bZ\\
E_8 &0 &0 &    0 &0 &0 &0 & 0 & 0
\end{array}
\]
\caption{Homotopy groups of Lie groups $\pi_d(G)$, $4\le d\le 11$.
 \label{table:homotopy}}
\end{table}

\bibliographystyle{ytphys}
\baselineskip=0.94\baselineskip
\bibliography{ref}

\end{document}